\documentstyle[epsf,prl,aps]{revtex}
\begin{document}
\twocolumn[\hsize\textwidth\columnwidth\hsize\csname@twocolumnfalse\endcsname
\preprint{\vbox{\hbox{November 1997} \hbox{IFP-746-UNC}  }}
\title{Constraints on free parameters of the simplest bilepton gauge model from 
the neutral kaon system mass difference}
\author{F. Pisano$^a$, J. A. Silva-Sobrinho$^b$ and M. D. 
Tonasse$^c$}
\address{$^a$Departamento de F\'\i sica, Universidade Federal do Paran\'a, 
81531-990 Curitiba, PR, Brazil\\
$^b$Instituto de F\'\i sica Te\'orica, Universidade Estadual Paulista, Rua 
Pamplona 145,\\
01405-900 S\~ao Paulo, SP, Brazil\\
$^c$Instituto de F\'\i sica, Universidade do Estado do Rio de Janeiro, Rua 
S\~ao Francisco Xavier 524\\
20550-013 Rio de Janeiro, RJ, Brazil} 
\maketitle

\begin{abstract}
We consider the contributions of the exotic quarks and gauge bosons to the 
mass difference between the short- and the long-lived neutral kaon states in 
SU(3)$_C\times$SU(3)$_L\times$U(1)$_N$ model. The lower bound $M_{Z^\prime} 
\sim$ 14 TeV is obtained for the extra neutral gauge boson ${Z^\prime}^0$. 
Ranges for values of one of the exotic quark masses and quark mixing 
parameters are also presented.
\end{abstract}
\vskip1pc]

\narrowtext
The $\Delta m = m_{K_L} - m_{K_S}$ mass difference between the long- and the 
short-lived kaon states was successfully used in a two 
generations standard model to predict the charmed quark mass\cite{GL74}. In 
the following years several authors have studied the $K^0-\overline{K^0}$ 
system in order to find constraints on parameters of new gauge theories such 
as gauge and scalar boson masses and mixing angles (see, for example, 
Refs.\cite{BL79,BB82,ST79}). The idea behind this is that, since the $c$ 
quark mass was predicted with a good precision, a possible new contribution 
to $\Delta m$ must be smaller than the result obtained from the two 
generations. 
\par
In this paper we go back to this subject in order to constrain a neutral 
gauge boson mass and quark mixing parameters imposed by the 3-3-1 
model\cite{PP92,FR92,PP97}. The 3-3-1 model is a gauge theory based on the 
SU(3)$_C\otimes$ SU(3)$_L\otimes$U(1)$_N$ semi-simple symmetry group. It has 
the interesting feature that the anomaly cancelation does not happen 
within each generation, as in the standard model, but only when the three 
generations are 
considered together. Thus, the number of families must be multiple of the 
color degrees of freedom and, as a consequence, the 3-3-1 model suggests a 
route towards the solution of the flavor question\cite{PP92,FR92,PP97,PI96}.\par
Let us summarize the most relevant points of the model. In the minimal 
particle content of Ref.~\cite{PP92} the left-handed quark fields transform under the SU(3)$_L$ group as the triplets
\begin{mathletters}
\begin{eqnarray}
Q_{1L} & = & \left(\begin{array}{ccc} 
         u_1 \\ d_{1\theta} \\ J_1
\end{array}\right)_L \sim ({\bf 3}, \frac{2}{3}), 
\label{quark1}\\
Q_{\alpha L} & = & \left(\begin{array}{ccc}
         J_{\alpha\phi} \\ u_\alpha \\ d_{\alpha\theta}
\end{array}\right)_L \sim ({\bf 3}^*, -\frac{1}{3})
\label{quark2}
\end{eqnarray}
\end{mathletters}
($\alpha$ = 2, 3), where 2/3 and $-$1/3 are the U(1)$_N$ charges. Each 
left-handed quark field has its right-handed counterpart transforming as a 
singlet of the SU(3)$_L$ group. In order to avoid anomalies one 
of the quark families must transform in a different way with respect to the 
two others. In Ref.\cite{FR92} the singled family is the third one, but this 
is not relevant here. The exotic quark $J_1$ carries 5/3 units of electric 
charge while $J_2$ and $J_3$ carry $-$4/3 each one of them. In the gauge 
sector the single charged $(V^{\pm})$ and the double charged 
$(U^{\pm\pm})$ vector bileptons\cite{CD97}, together with a new 
neutral gauge boson ${Z^\prime}^0$ complete the particle spectrum with the 
charged $W^\pm$ and the neutral $Z^0$ standard gauge bosons. At low energy 
the 3-3-1 model recovers the standard phenomenology\cite{PP97,LN94}.\par
An important property is that the bileptons can have a low energy scale. Low 
energy data constrain the vector bilepton masses to $M_X >$ 230 GeV 
$(X \equiv V^+, U^{++})$\cite{CF92}. Some authors have used the 
running of the coupling constant to impose upper bounds on 3-3-1 gauge boson 
masses\cite{SA93,LI94}. However, this procedure involves an arbitrary 
normalization of $N$\cite{FR92,PP97}. Usually this is done like in the 
standard model, although it is not mandatory. Recently these upper bounds 
were reexamined and it was found that the $M_{Z^\prime}$ mass has not upper 
bound, differently from previous calculations, while $M_X < 3.5$ 
TeV\cite{JJ97}.\par
The $\Delta m$ mass difference was already studied in the context of 
the 3-3-1 model at tree level (Fig. 1a) in order to put lower bound on the 
$M_{Z^\prime}$ mass\cite{DP94} and on mixing parameters\cite{LI94}, in the 
last case, taking into account an upper bound on $M_{Z^\prime}$. Here we 
apply the experimental lower limits and these reexamined upper bounds on the 
$M_X$ gauge boson masses in order to obtain the impositions of $\Delta m$ upon 
some of the 3-3-1 free parameters. Since the bileptons couple exotic to 
ordinary quarks, leading to additional contributions to the box diagram for 
the $K^0-\overline{K^0}$ transition (Fig. 1b), we combine the tree level with 
the box contribution.\par 
The charged current interactions for the quarks are given in Ref.\cite{PP92} 
and we can rewrite them as
\begin{eqnarray}
{\cal L} & = & -\frac{g}{2\sqrt{2}}\left[\overline{U}\gamma^\mu(1 - 
\gamma_5)V_{\rm CKM}DW^+_\mu +\right.\cr
 && \left. + \overline{U}\gamma^\mu(1 - 
\gamma_5)\zeta{\cal JV}_\mu +\right.\cr
&& \left. + \overline{D}\gamma^\mu(1 - 
\gamma_5)\xi{\cal JU}_\mu\right] + {\mbox {\rm H. c.}},
\end{eqnarray}
where
\begin{mathletters}
\begin{eqnarray}
U = \left(\begin{array}{c} 
         u \\ c \\ t
\end{array}\right), & \qquad &
D = \left(\begin{array}{c}
         d \\ s \\ b
\end{array}\right), \\
{\cal V}_\mu = \left(\begin{array}{c} 
         V^+_\mu \\ U^{--}_\mu \\ U^{--}_\mu 
\end{array}\right), & \qquad &
{\cal U}_\mu = \left(\begin{array}{c}
         U^{--}_\mu \\ V^+_\mu \\ V^+_\mu
\end{array}\right),
\end{eqnarray}
\end{mathletters}
and ${\cal J} = {\rm diag}\left(\begin{array}{ccc}J_1 & J_2 & J_3\end{array}\right)$. The $V_{\rm CKM}$ is 
the usual Ca\-bibbo-Kobayashi-Maskawa mixing matrix and $\zeta$ and $\xi$ are 
mixing matrices containing the new unknown parameters due to the presence of the 
exotic quarks. Here, unlike Eqs. (\ref{quark1},\ref{quark2}), we are working with the mass 
eigenstates.\par
\begin{figure}[h]
\begin{center}
\epsfxsize=2.8in
\epsfysize=10 true cm
\ \epsfbox{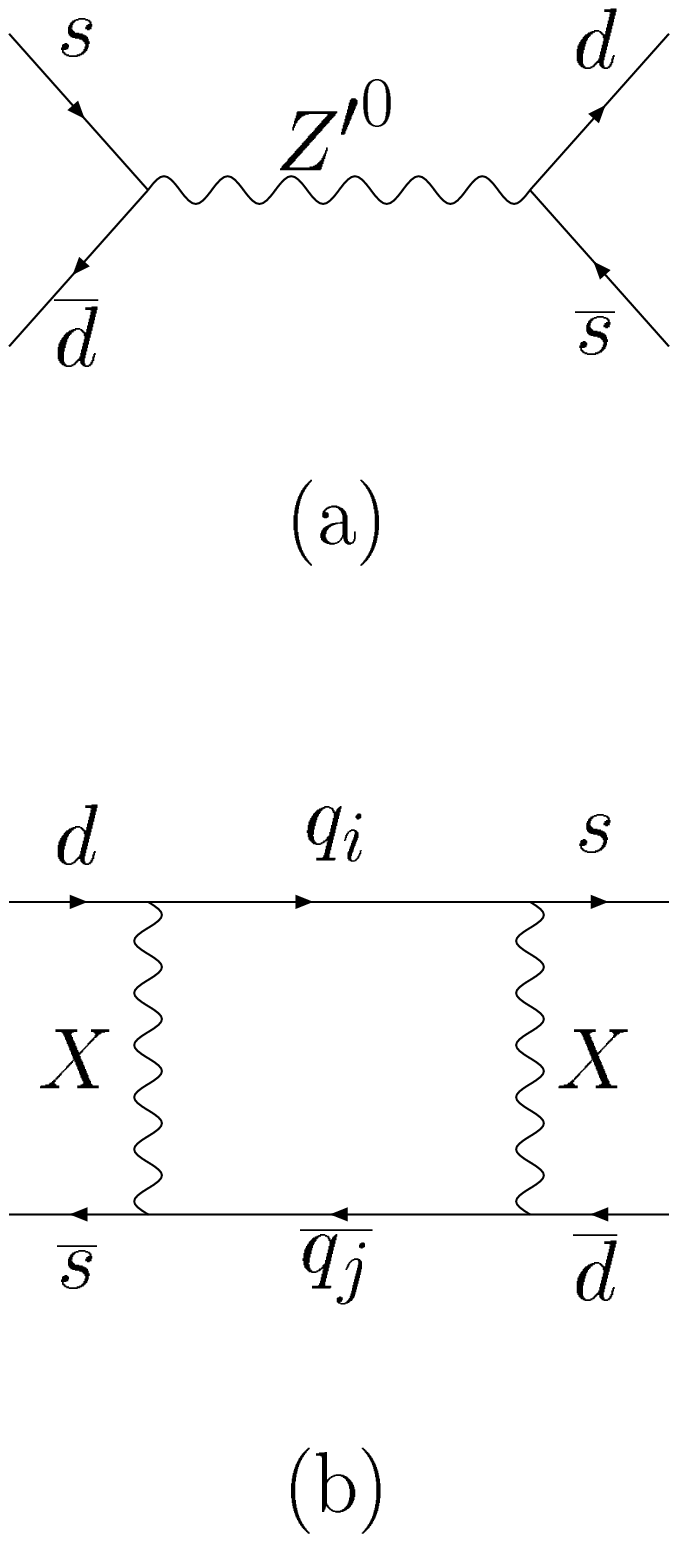}
\end{center}
\caption[]{$K^0-\overline{K^0}$ transition originating the $\Delta m$ mass difference at tree level (a) and the box diagram (b) in 3-3-1 model. The $q$'s are all quarks coupled by $X$ gauge bosons to $d$ and $s$.}
\label{fig:1}
\end{figure}
We are using the unitary gauge.  The box diagram for the $K^0-\overline{K^0}$ 
transition is represented in Fig. 1b, where in the standard model, $q_i = 
U_i$ $(i = 1, 2, 3)$ and $X$ is the $W^-$ boson. In the 3-3-1 
model, besides these, we have new contributions: for $i = 1$, $q_1 = J_1$ and 
$X = U^{--}$ while for $i = \alpha = 2, 3$, $q_\alpha = J_\alpha$ with $X = 
V^-$.  We do not consider here contributions of the scalar bileptons. This 
would be very complicated because of the proliferation of unknown parameters 
as mixing angles and scalar boson masses\cite{TO96}. However, because of the 
elusivity of the scalar particles, we expect that the new contributions of the 
3-3-1 model will be dominated by the gauge bosons.\par
Our calculation is standard\cite{GL74}. Neglecting long distance terms, we 
define ${\cal L}_X$ -- an effective Lagrangian corresponding to the $X$ gauge 
boson exchange in the box diagram -- as being the free quark amplitude 
$A_X(d\overline{s} \to \overline{d}s)$ and we evaluate the matrix 
element $\langle\overline{K^0}\vert-{\cal L}_X\vert K^0\rangle$. According 
the MIT bag model we can write\cite{ST79}
\begin{equation}
A_X = -0.7\left(\frac{G_FM_W^2}{\pi M_X}\right)^2{\cal 
O}_{sd}\sum_{i,j}\Gamma_i\Gamma_j\frac{f(a^X_{q_i}) - 
f(a^X_{q_j})}{a^X_{q_i} - a^X_{q_j}},
\label{ampl}
\end{equation} 
where the $\Gamma_i$ are mixing parameters given in the standard model by 
$V^*_{is}V_{id}$ and in the 3-3-1 model by $\xi_{is}^*\xi_{id}$. From the 
momentum integration we have
\begin{equation}
f(x) = \frac{1}{1 - x}\left(1 + \frac{x^2\ln{x}}{1 - x}\right).
\end{equation}
In Eq. (\ref{ampl}) $a_{q_i}^X = (m_{q_i}/M_X)^2$ is the square 
ratio of the $m_{q_i}$ quark and the $M_X$ gauge boson masses. We have defined 
${\cal O}_{sd} \equiv[\overline{v_s}\gamma_\mu(1 - 
\gamma_5)u_d]^2$. The sum in Eq. (\ref{ampl}) runs over all 
quarks coupled by the $X$ gauge boson. The $\Delta m_X$ mass 
difference due to the $X$ gauge boson contribution is given by
\begin{equation}
\Delta m_X = -2{\rm Re}\langle\overline{K^0}\vert{\cal L}_X\vert K^0\rangle 
\end{equation} 
with ${\cal L}_X \equiv A_X$. Taking into account the relation
\begin{equation}
\langle\overline{K^0}\vert{\cal O}_{sd}\vert K^0\rangle = \frac{4}{3}f_K^2m_K,
\end{equation}
where $m_K = 498$ MeV and $f_K = 161$ MeV are the kaon meson mass and its 
decay constant, respectively, the total mass difference is
\begin{equation}
\Delta m = 
\frac{8\eta 
m_K}{3}\left(\frac{f_KM_W}{s_WM_X}\right)^2\sum_{i,j}\Gamma_i\Gamma_j
\frac{f(a^X_{q_i}) - f(a^X_{q_j})}{a^X_{q_i} - 
a^X_{q_j}},
\label{dm}
\end{equation}
where $s_W = \sin{\theta_W}$, $\theta_W$ is the Weinberg weak mixing angle 
and we have included the leading order QCD correction factor $\eta = 
0.55$\cite{LI94,GW80}.\par
Despite the uncertainties undergone by the evaluation of the effective 
Lagrangian ${\cal L}_X$, the application of this method to well known 
processes tells us that the  procedure is reliable\cite{BB82}.\par
In order to get physical results from Eq. (\ref{dm}) we perform some 
hypotheses on the free parameters. Firstly we notice that the standard model 
does not give the whole $\Delta m$ mass difference. Considering only 
two quark generations we find $\Delta m_2 = 2\times10^{-15}$ GeV, while the 
experimental value is $\Delta m_{exp} = 3.491\times10^{-15}$ GeV\cite{Bet96}. 
The top quark contribution is negligible because of the smallness of the 
corresponding mixing parameters. Thus, if the 3-3-1 model is realized in 
Nature, we can expect that its pure contribution to $\Delta m$ is about 
$10^{-15}$ GeV.\par
Let us examine individual 3-3-1 contributions under the conservative 
hypothesis in which each one of them gives the total assumed 3-3-1 
counterpart to $\Delta m$, {\it i. e.}, $10^{-15}$ GeV. We begin analyzing the contribution of the $U$
bilepton and the $J_1$ quark. By convenience we assume for the $\xi$ mixing matrix the same parametrization advocated by the Particle Data Group for 
$V_{CKM}$\cite{Bet96}. In this case, since we choose all the mixing angles in the first quadrant, the parameter $\xi^*_{1d}\xi_{1s}$ is positive. We do not 
consider here CP violating phases in the mixing matrices. In Fig. 2 we plot this mixing parameter as function of $M_U$, according Eq. (\ref{dm}), where we are using the bounds 
$M_U > 230$ GeV\cite{CF92} and $M_U < 3500$ GeV\cite{JJ97}. Hence, we can see 
that $0.014 \lesssim \xi^*_{1d}\xi_{1s} \lesssim 0.35$. Physically this range 
for the mixing angles implies $m_{J_1} \lesssim M_U$. In order to apply the  
lower limit on the mixing parameter for obtaining a lower bound for 
$M_{Z^\prime}$, we use the result of Ref.\cite{DP94} for the $Z^\prime$ 
contribution to $\Delta m$, but taking into account that we are assuming that 
the maximal contribution to $\Delta m$ is $\sim 10^{-15}$ GeV (not the whole 
experimental value). We introduce also the leading order QCD correction 
factor $\eta = 0.55$. This leads to $M_{Z^\prime} \sim 
1.03 \times 
10^3\left[{\rm Re}\left({V^D_L}_{11}^*{V^D_L}_{12}\right)\right]^{1/2}$ TeV, 
where $V_L^D$, the mixing matrix relating the symmetry left-handed quark 
states carrying -1/3 units of electric charge $\left(D_L^\prime\right)$ with the 
physical ones $\left(D_L\right)$, is defined by $D^\prime_L = V_L^DD_L$. 
Since only the $J_1$ quark carries 5/3 units of electric charge it no mix. 
Therefore, the mixing parameter ${V_L^D}_{11}^*{V_L^D}_{12}$ is the same as 
$\xi_{1s}^*\xi_{1d}$, whose lower bound provides $M_Z^\prime \sim 14.4$ TeV.\par
A more rigorous analysis, considering the contributions of diagrams 
exchanging the single charged $V$ bilepton, $J_2$ and $J_3$ exotic quarks, is 
complicated since these two quarks can mix and the sign of the mixing 
parameters is not defined according to our parametrization of the mixing 
matrices and the choice of the mixing angles. However it is not expected an 
appreciable difference in the results.\par
\begin{figure}[h]
\begin{center}
\epsfxsize=2.8in
\epsfysize=2.3in
\ \epsfbox{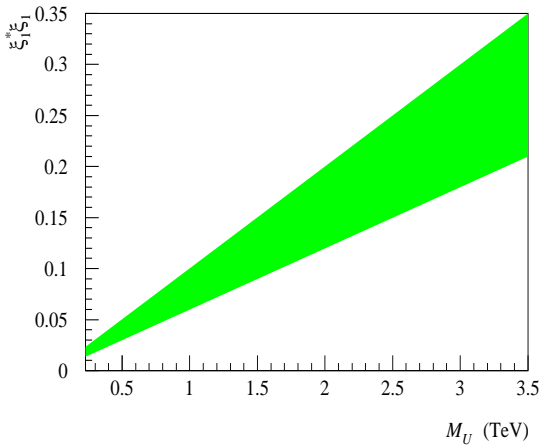}
\end{center}
\caption[]{Bounds on quark mixing parameters from $\Delta m$ mass 
difference as function of the double charged bilepton mass. The dark region 
represents the allowed values for $\xi_{1d}^*\xi_{1s}$.}
\label{fig:2}
\end{figure}
In this letter we have studied the implications of the $\Delta m$ 
mass difference on free parameters of the 3-3-1 model. However, differently 
from previous calculations\cite{LI94,DP94}, here we are taking into account 
the contribution of the box diagram exchanging the double charged $U^{--}$ 
bilepton. Our results differ from the previous ones because we have combined 
the analysis for the tree level and the box diagram. Therefore, if the 3-3-1 
Higgs contribution to $\Delta m$ is not important we can assume the value we 
have estimated for $M_{Z^\prime}$ as an approximate lower bound ({\it i. e.}, 
$M_{Z^\prime} \stackrel{>}{\sim} 14$ TeV). We stress that this bound does not 
depend on the aforementioned arbitrary normalization of $N$. The crucial 
parameter for this value is the experimental lower bound on the mass of the 
$U^{--}$ bilepton gauge boson.
\acknowledgements
We would like to thank Dr. M. C. Tijero for reading the manuscript and the 
Instituto de F\'\i sica Te\'orica -- UNESP for use of its facilities. The 
work of M. D. T. was supported by the Funda\c c\~ao de Amparo \`a Pesquisa no 
Estado do Rio de Janeiro (proc. n$^{_{\d{\rm o}}}$ E-26/150.338/97).

\end{document}